\documentclass[pdflatex,sn-mathphys-num]{sn-jnl}

\usepackage{graphicx}%
\usepackage{multirow}%
\usepackage{amsmath,amssymb,amsfonts}%
\usepackage{amsthm}%
\usepackage{mathrsfs}%
\usepackage[title]{appendix}%
\usepackage{xcolor}%
\usepackage{textcomp}%
\usepackage{manyfoot}%
\usepackage{booktabs}%
\DeclareUnicodeCharacter{207B}{\textsuperscript{-}}
\usepackage{algorithm}%
\usepackage{algorithmicx}%
\usepackage{algpseudocode}%
\usepackage{listings}%

\theoremstyle{thmstyleone}%
\theoremstyle{thmstyletwo}%

\theoremstyle{thmstylethree}%

\begin{document}

\title[Article Title]{Cast3: Translating numerical weather prediction principles into data-driven forecasting}


\author[1]{\fnm{Congyi} \sur{Nai}}\email{ncy17@tsinghua.org.cn}

\author[1]{\fnm{Baoxiang} \sur{Pan}}\email{panbaoxiang@lasg.iap.ac.cn}

\author[2]{\fnm{Yuan} \sur{Liang}}\email{yuan.liang@smail.nju.edu.cn}

\author*[1]{\fnm{Xi} \sur{Chen}}\email{chenxi@lasg.iap.ac.cn}


\affil[1]{\orgdiv{State Key Laboratory of Earth System Numerical Modeling and Application}, \orgdiv{Institute of Atmospheric Physics}, \orgname{Chinese Academy of Science}, \city{Beijing}, \country{China}}

\affil[2]{\orgname{TianJi Weather Science and Technology Company}, \city{Beijing}, \country{China}}


\abstract{Data-driven weather models have made rapid advances in recent years, reaching and in some metrics surpassing the large-scale forecast skill of operational numerical weather prediction. This progress, however, has been built almost entirely on the reanalysis data that NWP produced, while the methodological knowledge that the NWP community distilled over decades of multi-scale atmospheric modelling remains largely unused. Here we present Cast3, a generative forecasting framework that systematically absorbs NWP meta-knowledge to close this gap. Cast3 operates on variable-resolution cubed-sphere grids for scale-aware representation and constructs structurally diverse super-ensembles that sample the complementary biases of different grid discretizations, delivering state-of-the-art ensemble prediction. It further introduces generative nudging, a posterior-sampling strategy that distils the collective information of the full ensemble into a single forecast possessing both the large-scale accuracy of the ensemble mean and the mesoscale realism of a high-resolution member. Evaluated across synoptic-scale skill, spectral fidelity, station-level surface verification, and tropical cyclone prediction, Cast3 outperforms established deterministic and generative baselines across various dimensions. More broadly, these results demonstrate that the design principles embedded in computational atmospheric science offer a rich and largely untapped foundation for the next generation of data-driven Earth system modelling. Real-time forecasts are publicly available at \url{https://project.iap.ac.cn/AI/index.html}.}

\maketitle

\section{Introduction}\label{sec1}

Weather forecasting is, at its core, a problem of evolving a high-dimensional, chaotic field forward in time. The atmosphere is governed by the Navier-Stokes equations coupled to thermodynamic and radiative transfer relations on a rotating sphere. These equations are deterministic in principle but chaotic in practice: small perturbations in initial conditions amplify exponentially, imposing a finite predictability horizon beyond which point forecasts lose meaning \cite{lorenz_deterministic_1963}. One practical consequence shapes everything that follows: large-scale circulation patterns carry a predictable signal that can be extracted and extended in time, while fine-scale weather structures are governed by uncertainty that must be represented rather than suppressed.

Since the first numerical weather prediction (NWP) by \citet{charney_numerical_1950}, nearly eight decades of sustained effort have produced operational forecasting systems of remarkable skill. Modern NWP systems routinely produce useful guidance at five to ten days for large-scale circulation patterns \cite{bauer_quiet_2015}, yet operational global models still cannot routinely resolve the mesoscale processes that govern the most impactful weather events. What this effort has produced, however, extends well beyond forecast data. The NWP community has accumulated a body of methodological knowledge distilled from confronting precisely the multi-scale challenge described above: how to represent atmospheric states on the sphere \cite{sadourny_conservative_1972,adcroft_implementation_2004,marras_simulation_2015,putman_finite-volume_2007,chen_lmars_2021}, how to treat the interaction between predictable large-scale flow and uncertain fine-scale structures \cite{waldron_sensitivity_1996,von_storch_spectral_2000}, and how to scale computation across thousands of processors while preserving physical continuity \cite{michalakes_avec_2015,national_energy_research_scientific_computing_center_u.s._avec_2016}. This meta-knowledge is not about the data itself but about how to represent, compute, and reason about atmospheric fields across scales.

The past several years have witnessed a striking disruption. Data-driven models trained on reanalysis archives have reached and in some metrics surpassed the large-scale forecast skill of operational NWP, at a fraction of the computational cost \cite{bi2023accurate,chen2023fuxi,lam_learning_2023,lang_aifs_2024,alet2025skillful,bonev2025fourcastnet,chen2025operational,nai2025boosting,price_probabilistic_2025,kossaifi2026demystifying,lang2026aifs}. Yet this rapid progress is approaching a plateau. A growing number of models now cluster around similar performance levels, built from recognizably similar components and trained on the same reanalysis corpus \cite{bonavita_limitations_2024}. The most conspicuous shared deficiency is that forecasts with high deterministic skill are spatially smooth \cite{rasp2024weatherbench,price_probabilistic_2025,polichtchouk2026hybrid}, lacking the fine-scale structure critical for high-impact weather guidance, while methods that preserve mesoscale texture sacrifice the deterministic accuracy that decision-makers require \cite{price_probabilistic_2025}. Accurate large-scale prediction and realistic fine-scale detail have not yet been achieved together in a single data-driven forecast \cite{husain2025leveraging,polichtchouk2026hybrid}.

We argue that a key reason for this impasse is that the machine-learning weather prediction community has drawn heavily on the data \cite{hersbach2020era5} accumulated by numerical weather prediction but has not yet systematically exploited the methodological knowledge that the NWP community distilled over decades. This meta-knowledge encompasses at least three categories of hard-won insight. First, representation: the choice of computational mesh determines which physical structures can be faithfully resolved and how errors project onto different scales. NWP developed variable-resolution meshes such as stretched cubed-sphere grids \cite{harris_high-resolution_2016} concentrate degrees of freedom where forecast skill is most resolution-sensitive, yet data-driven models have overwhelmingly operated on uniform grids. Second, scale interaction and ensemble reasoning: spectral nudging constrains the large-scale circulation toward a trusted reference while allowing small-scale dynamics to evolve freely \cite{waldron_sensitivity_1996,von_storch_spectral_2000}; the ensemble mean extracts a predictable signal that no single member can isolate, yet destroys fine-scale structure in the process \cite{toth_ensemble_1997}. How to recombine this information into a single forecast that is simultaneously skillful at large scales and realistic at small scales remains an open question even within NWP. Third, computation: NWP can achieve global coverage at high resolution through domain decomposition on cubed-sphere grids, where the six faces of the cube are partitioned across processors with boundary information exchanged at every time step \cite{putman_finite-volume_2007}. This provides a natural blueprint for scaling data-driven models beyond single-accelerator memory limits.

Here we present Cast3, a generative weather forecasting framework designed to translate these categories of NWP meta-knowledge into the architecture, training, and inference of a data-driven system. Cast3 operates natively on stretched cubed-sphere grids, the same variable-resolution topology used in modern dynamical cores such as FV3 \cite{putman_finite-volume_2007}, concentrating generative capacity \cite{ho2020denoising,song2020score} where fine-scale skill is most needed. Training independent diffusion models on distinct grid configurations and sampling multiple realizations from each produces a structurally diverse super-ensemble whose members differ not only in their stochastic trajectories but in their regional emphasis and discretization errors, an ensemble design philosophy borrowed directly from multi-model NWP practice. We further introduce \textit{generative nudging}, a posterior-sampling strategy \cite{chung2022diffusion} that embeds scale-selective guidance into the diffusion denoising process. At each reverse-diffusion step a gradient-based correction nudges the sample toward the ensemble mean at trustworthy scales, while the learned prior generates fine-scale structures consistent with the large-scale constraint, directly analogous to lead-time-dependent spectral nudging in regional NWP. The result is a single Cast3 forecast that inherits the deterministic skill of the ensemble mean and the mesoscale realism of an individual high-resolution member. The cubed-sphere topology further supports spatial parallelism through differentiable halo exchange \cite{chen_lmars_2021}, enabling inference and, prospectively, training at resolutions beyond single-device memory.

We evaluate Cast3 across global 500 hPa geopotential height, 200 hPa kinetic energy spectra, station-level two-metre temperature verification over China, and tropical cyclone track and structure over the Western North Pacific. Cast3 outperforms established deterministic and generative baselines across all four evaluation dimensions. More broadly, Cast3 illustrates that the meta-knowledge embedded in mature computational science disciplines constitutes a rich and largely untapped design space for data-driven modelling.

\section{Cast3 forecast system}\label{sec2}

\begin{figure}[H]
\centering
\includegraphics[width=\textwidth]{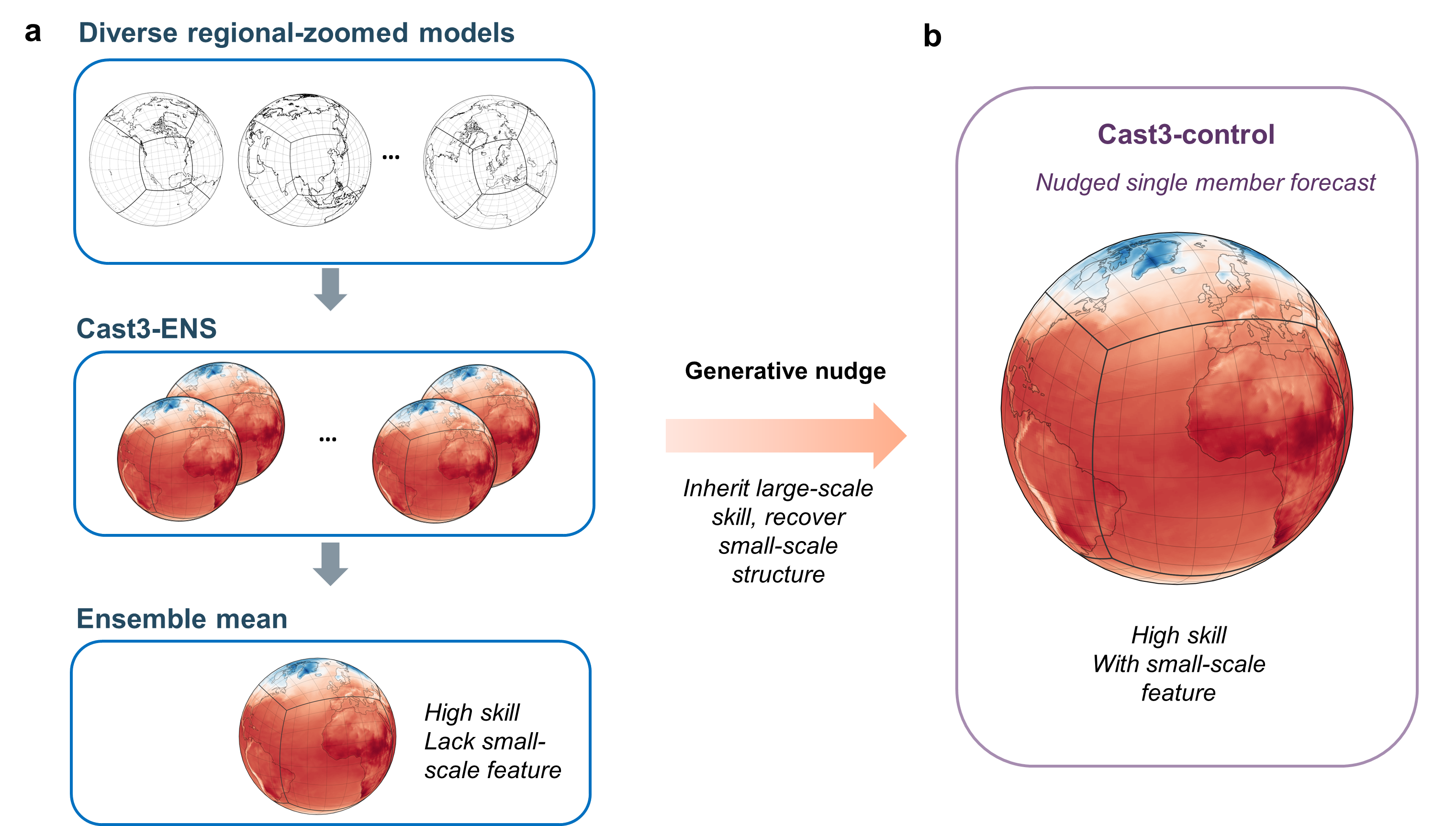}
\caption{\textbf{Schematic of the Cast3 framework.}
\textbf{a,}~The diffusion model generates Cast3-ENS by sampling $K$ ensemble members from independent noise initializations across multiple stretched cubed-sphere grid configurations (top: example grid with high effective resolution over the target region). The super-ensemble mean $\bar{X}$ is computed by aggregating all members.
\textbf{b,}~Generative nudging applies diffusion posterior sampling (DPS) conditioned on $\bar{X}$ to produce Cast3-Control. Large-scale wavenumbers from $\bar{X}$ constrain the sampling, while small-scale features are generated stochastically. The constraint wavelength varies with lead time.}
\label{fig:framework}
\end{figure}

A schematic of the Cast3 framework is shown in Fig.~\ref{fig:framework}. The system comprises two stages operating within a unified diffusion-based probabilistic framework: ensemble generation and generative nudging, connected through the ensemble mean.

The foundation of Cast3 is a conditional diffusion model that operates on stretched cubed-sphere grids (Fig.~\ref{fig:framework}a). By shifting the high-resolution face of the cube toward a chosen target region, each grid configuration resolves local dynamics at an effective high resolution while maintaining global context at coarser scales. Its structured topology is naturally compatible with standard convolutional operations, enabling efficient training and inference.

Centering the high-resolution face on different geographical regions yields a family of distinct grid configurations. Training an independent diffusion model on each configuration and sampling multiple realizations from each produces a structurally diverse super-ensemble (Cast3-ENS) whose members differ in both their regional emphasis and their stochastic trajectories, jointly capturing mesh-configuration, model, and initial-condition uncertainty (see details in Methods). Aggregating these members into an ensemble mean preserves the skillful large-scale circulation but inevitably smooths out fine-scale features.

To recover these features, we introduce generative nudging (Fig.~\ref{fig:framework}b), which applies diffusion posterior sampling (DPS) conditioned on the ensemble mean using any single region-focused model from the first stage. At each denoising step, the sample is guided toward the ensemble mean through a gradient-based correction, while the diffusion process simultaneously generates fine-scale structures that feed back onto the large-scale flow. This interplay produces a single realization, Cast3-Control, that inherits the synoptic-scale skill of the mean while exhibiting physically realistic mesoscale detail. As lead time increases, predictability decreases and ensemble members progressively diverge, agreeing only at increasingly large scales \cite{roberts_assessing_2008,surcel_study_2015}. The ensemble mean therefore loses effective resolution over time, as small-scale features with inconsistent phases cancel upon averaging. Generative nudging adapts to this by guiding only those scales at which the mean preserves physically meaningful information, with the precise cutoff determined by the effective resolution of the ensemble mean at each lead time (see details in Methods). Crucially, because this stage operates independently on the high-resolution domain, it can be trained directly on km-scale numerical simulation data, allowing the system to learn physical structures and variability that are absent from coarser reanalysis products.

\section{Large-scale skill and spectral realism}\label{sec3}

We benchmark Cast3 against the leading operational and machine-learning baselines: GenCast, the ECMWF ensemble (IFS-ENS), and the high-resolution deterministic forecast IFS-HRES (hereafter IFS-Control). Verification is performed over 38 cases spanning June--December 2023, using the 500~hPa geopotential height anomaly correlation coefficient (Z500 ACC) as a measure of synoptic-scale predictive skill and the 200~hPa kinetic energy spectrum as a diagnostic of dynamical realism across scales.

\begin{figure}[H]
\centering
\includegraphics[width=\textwidth]{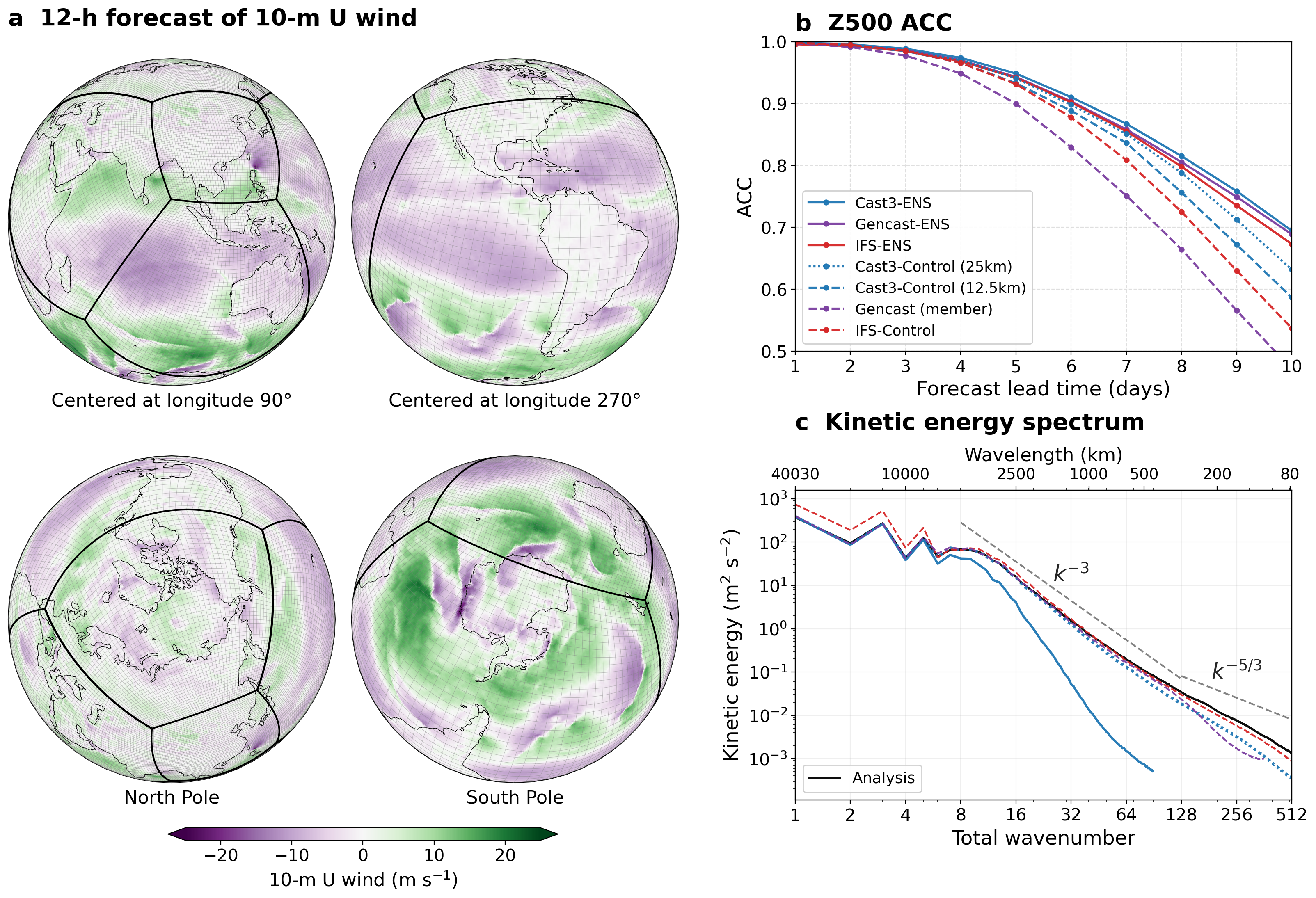}
\caption{\textbf{Large-scale skill and spectral realism of Cast3.}
\textbf{a,}~A 12~h forecast of 10~m zonal wind ($\mathrm{m\,s^{-1}}$) from a single Cast3-ENS member, rendered on the native cubed-sphere mesh and viewed from four perspectives.
\textbf{b,}~Global 500~hPa geopotential height anomaly correlation coefficient (Z500 ACC) as a function of forecast lead time. Solid lines denote ensemble means; dashed lines denote individual members or deterministic forecasts.
\textbf{c,}~Kinetic energy spectrum at 200~hPa for day-10 forecasts. Reference slopes of $k^{-3}$ and $k^{-5/3}$ are shown in grey dashes.}
\label{fig:benchmark}
\end{figure}

Figure~\ref{fig:benchmark}a presents the native output of a single Cast3-ENS member: a 12~h forecast of 10~m zonal wind shown from four viewpoints. The forecast is rendered directly on the model mesh, exposing the six-panel cubed-sphere topology and the smooth refinement from a 25~km quasi-uniform global background to a 12.5~km target region over East Asia. The field remains continuous across all panel boundaries and through the refinement zone, with no blending or interpolation artifacts, illustrating the seamless global-to-regional formulation that underpins every evaluation that follows.

Figure~\ref{fig:benchmark}b shows global Z500 ACC as a function of lead time. Cast3-ENS attains the highest skill at every lead time, reaching 0.694 at day~10 compared with 0.689 for GenCast-ENS and 0.673 for IFS-ENS, with its advantage over both baselines widening at longer horizons. As expected, for all three systems the ensemble mean substantially outperforms individual members in large-scale skill, though at the cost of suppressing fine-scale energy as revealed by the kinetic energy spectrum (Fig.~\ref{fig:benchmark}c).

Cast3-Control is designed to overcome this trade-off. Through generative nudging, a single realization is guided toward the Cast3 ensemble mean, inheriting its synoptic-scale pattern while preserving intrinsic mesoscale structure (see Methods). We evaluate two configurations: a 25~km global Cast3-Control and a 12.5~km variant whose diffusion model is trained on IFS forecasts rather than reanalysis, which limits its standalone skill (see Methods). At day~10, the 25~km and 12.5~km configurations attain global Z500 ACC values of 0.632 and 0.586 respectively, both substantially exceeding the deterministic baselines of IFS-Control (0.537) and GenCast (0.471). The 12.5~km configuration is particularly revealing: although its training target is the IFS forecast itself, which in principle bounds its achievable skill by that of a single IFS member, generative nudging lifts it well above IFS-Control. This demonstrates that, by exploiting the complementary information content of multiple ensemble members and data sources, Cast3-Control transcends the skill ceiling imposed by any single one of them.

Figure~\ref{fig:benchmark}c presents the 200~hPa kinetic energy spectrum at day~10. The Cast3 ensemble mean loses power from synoptic wavenumbers onward and decays rapidly into the mesoscale, the expected signature of phase cancellation upon averaging. Cast3-Control, by contrast, tracks the analysis spectrum closely, following the canonical $k^{-3}$ slope throughout the synoptic range and recovering a $k^{-5/3}$ segment at mesoscale wavenumbers, indicating that genuine mesoscale dynamics are retained rather than reconstructed from smoothed fields. GenCast exhibits pronounced spectral damping at small scales even in its deterministic mode, plausibly reflecting the resolution ceiling of its ERA5 training data. Together, these results demonstrate that Cast3 simultaneously achieves two objectives conventionally in tension: the synoptic-scale predictability of an ensemble mean and the mesoscale realism of an individual member. Generative nudging reconciles the two within a single forecast, delivering leading large-scale skill alongside a physically consistent energy spectrum across the full range of resolved scales.

\section{Fine-scale skill at station level}\label{sec4}

Gridded verification alone is insufficient to assess the fidelity of fine-scale features; in-situ station observations remain the most direct benchmark for evaluating whether a forecast captures local meteorological details. Among surface variables, two-metre temperature (T2m) is particularly demanding because it is strongly modulated by terrain. Valleys, basins, and elevated plateaux impose sharp spatial gradients that only a forecast with adequate small-scale structure can reproduce. We therefore use T2m observations from more than 2,400 national stations across China to evaluate Cast3-Control against the ensemble mean (Cast3-ENS), IFS-Control, and a single GenCast member over lead times of one to seven days.

\begin{figure}[H]
\centering
\includegraphics[width=\textwidth]{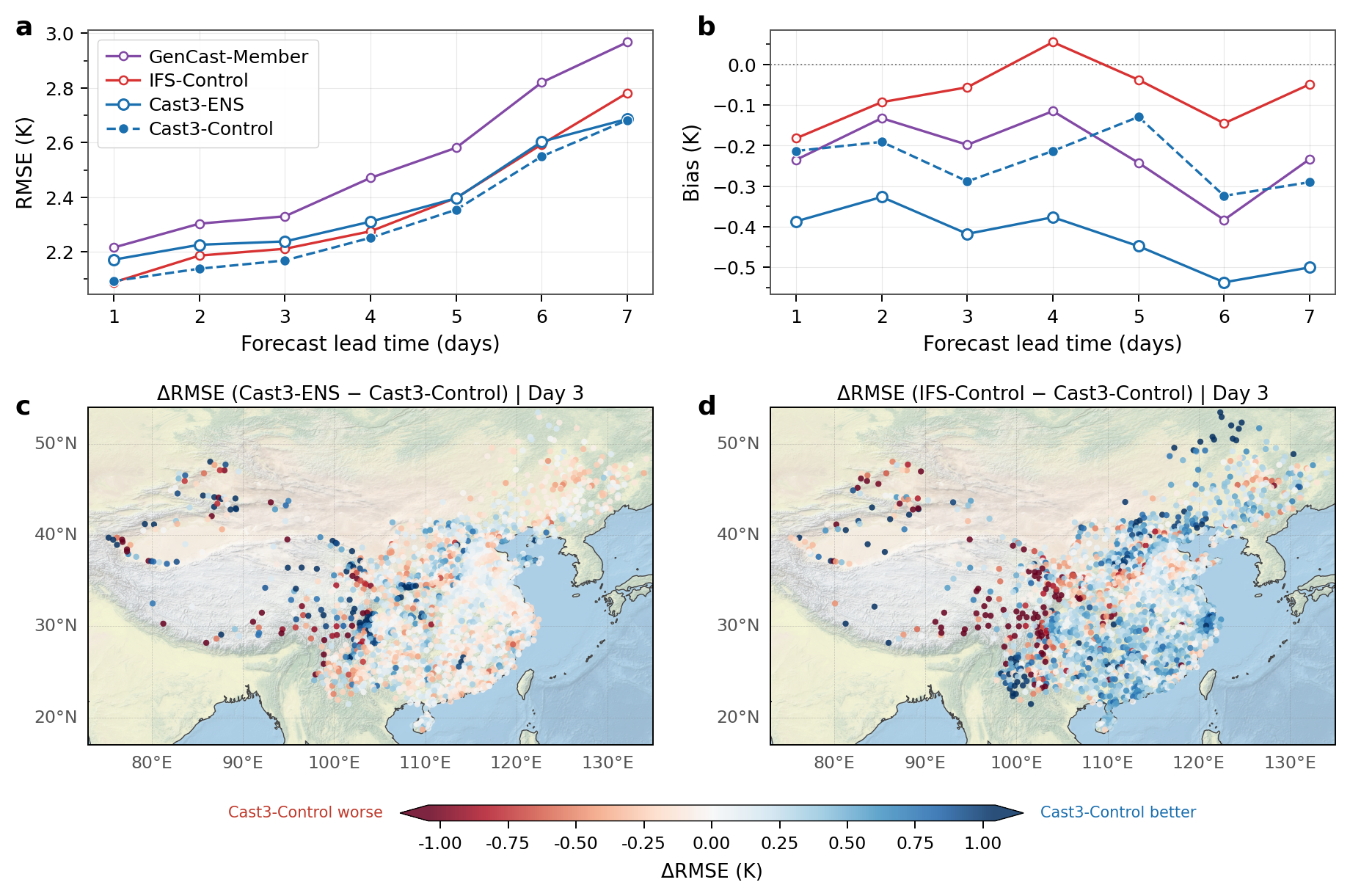}
\caption{Station-based verification of two-metre temperature over China.
\textbf{a,}~Root-mean-square error (RMSE) of T2m forecasts verified against more than 2,168 national stations, shown as a function of lead time for Cast3-Control, Cast3-ENS (ensemble mean), IFS-Control, and a single GenCast member.
\textbf{b,}~Corresponding mean bias (forecast minus observation).
\textbf{c,}~Station-level $\Delta$RMSE at day~3 between Cast3-ENS and Cast3-Control; positive values (blue) indicate Cast3-Control outperforms the ensemble mean.
\textbf{d,}~As in \textbf{c} but comparing IFS-Control against Cast3-Control; positive values (blue) indicate Cast3-Control outperforms IFS-Control.}
\label{fig:station}
\end{figure}

Figure~\ref{fig:station}a presents RMSE as a function of forecast lead time. Cast3-Control achieves the lowest RMSE at every lead time from day~1 through day~7, demonstrating a consistent advantage over all competing products. Notably, it outperforms IFS-Control despite the latter operating at roughly 9~km resolution, a grid spacing four to five times finer than that of the underlying ensemble. A single GenCast member, while retaining small-scale texture, exhibits the highest RMSE across all lead times, reflecting the limited deterministic skill of an individual stochastic realization. The ensemble mean tracks Cast3-Control closely but sits systematically above it, confirming that generative nudging adds genuine small-scale information rather than merely replicating the large-scale envelope. In this sense, Cast3-Control combines the deterministic skill of the ensemble mean with the fine-scale detail that averaging suppresses.

The bias characteristics shown in Figure~\ref{fig:station}b further clarify the physical origin of this improvement. All products exhibit a negative (cold) bias, but the ensemble mean carries the largest magnitude throughout the forecast range, reflecting the well-known tendency of ensemble averaging to suppress local extremes. Because each member resolves terrain-induced temperature contrasts slightly differently, the mean smears these contrasts and produces systematically cooler estimates at station locations, particularly in topographically complex regions. Cast3-Control substantially corrects this cold bias at all lead times, recovering much of the fine-scale variability that averaging erases, while IFS-Control maintains a relatively steady moderate bias that grows only slowly with lead time.

Figures~\ref{fig:station}c and~\ref{fig:station}d decompose the RMSE improvement spatially by mapping station-level $\Delta$RMSE at day~3, where positive values (blue) indicate that Cast3-Control outperforms the comparator. Comparing Cast3-Control against the ensemble mean (Fig.~\ref{fig:station}c), widespread improvement appears over the complex terrain of the Yunnan--Guizhou Plateau, the eastern flanks of the Tibetan Plateau, and the mountainous basins of Xinjiang. These are precisely the regions where small-scale topographic forcing is strongest and where the smoothing inherent in ensemble averaging is most damaging. This pattern confirms that generative nudging preferentially restores skill in areas that demand fine-scale terrain representation. Over the eastern plains, the improvement is more modest, consistent with the ensemble mean already performing adequately where the terrain is smooth and large-scale patterns dominate.

The comparison with IFS-Control (Fig.~\ref{fig:station}d) reveals a complementary picture. Over the Tibetan Plateau and other regions of steep orography, IFS-Control benefits from its substantially higher native resolution and shows lower RMSE than Cast3-Control, an expected result given that a 9~km grid can explicitly resolve terrain features that remain sub-grid at the ensemble's coarser resolution. However, across the densely observed plains of eastern and central China, Cast3-Control consistently outperforms IFS-Control, evidenced by a broad swath of improvement extending from the North China Plain through the middle and lower Yangtze basin. This advantage arises because Cast3-Control inherits the superior large-scale circulation of the ensemble mean, refined by aggregating across multiple plausible atmospheric trajectories, and thus starts from a more skillful synoptic-scale background. Where terrain complexity is low and forecast quality hinges primarily on the accuracy of the large-scale flow, this inherited advantage translates directly into lower station-level RMSE.

\section{Tropical cyclone track and structure}

Tropical cyclone forecasting demands both accurate tracks for evacuation planning and realistic fine-scale structure for assessing local wind and precipitation hazards, making it a critical test of any forecast system's ability to deliver large-scale skill and mesoscale realism simultaneously. We evaluate Cast3-Control using track errors aggregated over 13 tropical cyclones that affected the Western North Pacific after 15 July 2023, and examine the fine-scale structure in detail through a representative case initialized at 0000~UTC 20 July 2023.

\begin{figure}[H]
\centering
\includegraphics[width=\textwidth]{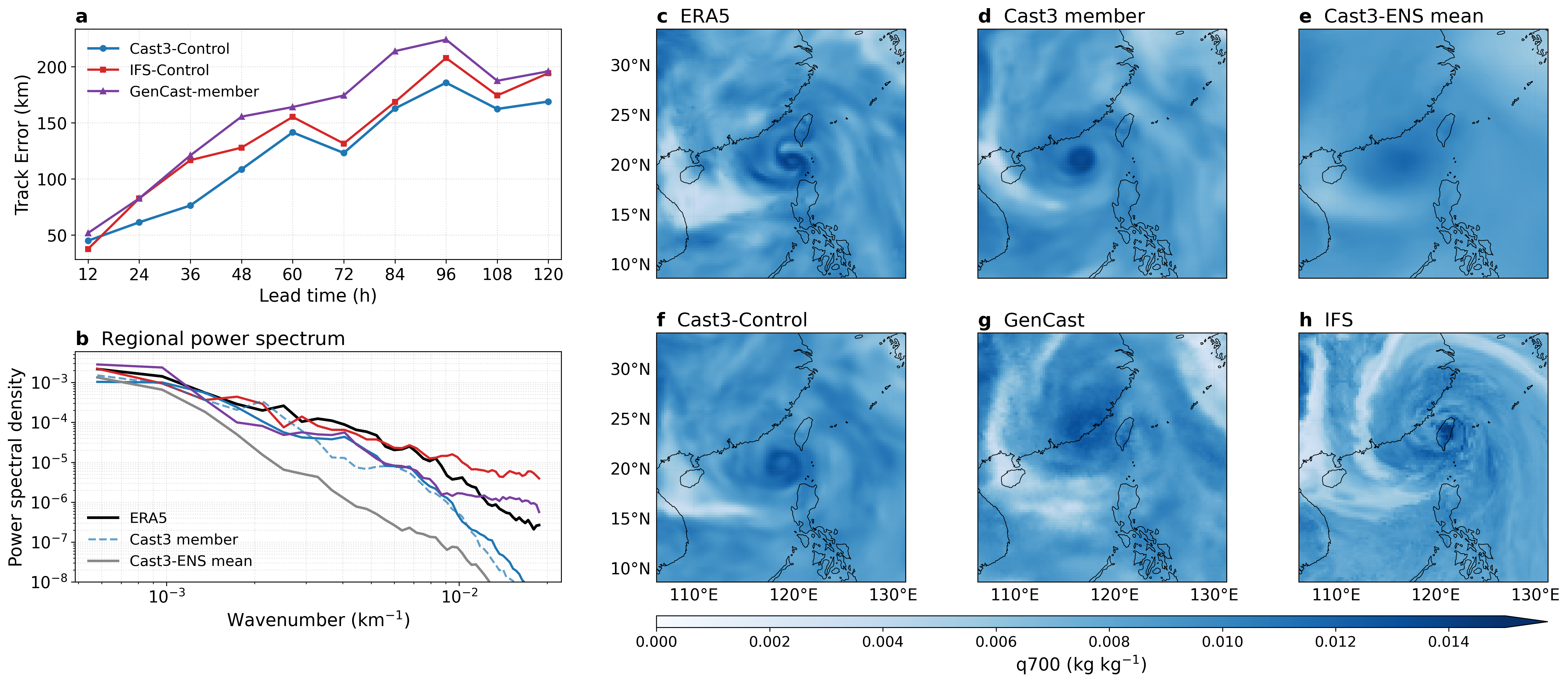}
\caption{\textbf{Tropical cyclone track skill and fine-scale structure.}
\textbf{a,}~Track error for Cast3-Control, IFS-Control, and a single GenCast member, aggregated over 13 tropical cyclones affecting the western North Pacific after 15 July 2023.
\textbf{b,}~Regional power spectral density of 700~hPa specific humidity over the typhoon domain for Typhoon Doksuri (initialized 0000~UTC 20 July 2023, verified at day~7).
\textbf{c--h,}~Day-7 forecasts of 700~hPa specific humidity ($q_{700}$; kg\,kg$^{-1}$) from the same initialization, shown alongside the ERA5 reanalysis valid at 0000~UTC 27 July 2023:
\textbf{c,}~ERA5 reanalysis;
\textbf{d,}~a single Cast3-ENS member;
\textbf{e,}~Cast3 ensemble mean;
\textbf{f,}~Cast3-Control;
\textbf{g,}~GenCast member;
\textbf{h,}~IFS-Control.}
\label{fig:typhoon}
\end{figure}

Figure~\ref{fig:typhoon}a shows track error as a function of lead time, computed as the great-circle distance between the forecast and observed cyclone positions. Cast3-Control achieves the lowest track error at nearly every lead time out to 120~h, reflecting the synoptic-scale accuracy inherited from the ensemble mean. IFS-Control is competitive at short lead times but degrades more rapidly beyond 72~h, while GenCast exhibits the largest track errors throughout the forecast range.

The regional power spectrum (Fig.~\ref{fig:typhoon}b) reveals how much energy each product retains at different spatial scales for the representative case. The Cast3 ensemble mean loses energy rapidly below wavelengths of a few hundred kilometres, the expected consequence of phase cancellation upon averaging. A single Cast3-ENS member and Cast3-Control both recover small-scale energy close to the ERA5 reference, confirming that generative nudging restores fine-scale variability without degrading the large-scale flow. GenCast exhibits an artificial upturn in energy at the smallest resolved wavelengths, a known artefact of diffusion-based models operating in spectral space \cite{kim2026spectral}, likely compounded by its formulation on a higher-dimensional state space that makes convergence at small scales more difficult. IFS, as a high-resolution physical model, naturally carries substantial energy at small scales, though its spectral slope departs from the reanalysis at the finest wavelengths.

Figures~\ref{fig:typhoon}c--h translate these spectral differences into physical space through 700~hPa specific humidity ($q_{700}$) fields, comparing day-7 forecasts against the ERA5 reanalysis (Fig.~\ref{fig:typhoon}c). The reanalysis shows a well-defined cyclonic moisture envelope with clearly delineated spiral banding and a compact moist core. A single Cast3-ENS member (Fig.~\ref{fig:typhoon}d) captures the cyclonic circulation and moisture asymmetry but with stochastic variability in the fine-scale features, while the ensemble mean (Fig.~\ref{fig:typhoon}e) retains only the broadest moisture signature, with all spiral structure and inner-core gradients smoothed beyond recognition. Cast3-Control (Fig.~\ref{fig:typhoon}f) recovers a coherent cyclonic moisture pattern with sharper gradients and discernible banding structure, representing a marked improvement over the ensemble mean while preserving the correct large-scale position inherited from it. GenCast (Fig.~\ref{fig:typhoon}g) produces a diffuse, nearly featureless moisture field that lacks organized convective structure, consistent with its energy deficit at intermediate scales. IFS-Control (Fig.~\ref{fig:typhoon}h) exhibits the most detailed spiral banding and inner-core structure, as expected from a high-resolution deterministic physical model, though its track error at this lead time places the cyclone in a noticeably different position.

Taken together, the tropical cyclone evaluation demonstrates the central advantage of generative nudging: Cast3-Control inherits the superior track accuracy of the ensemble mean, as shown by the consistently low position errors in Fig.~\ref{fig:typhoon}a, while simultaneously recovering the mesoscale moisture and dynamical structure that averaging destroys, as evidenced by the restored spectral energy in Fig.~\ref{fig:typhoon}b and the physically coherent cyclone structure in Fig.~\ref{fig:typhoon}f.

\section{Discussion}

Cast3 demonstrates that the large-scale skill advantage of data-driven ensemble forecasting can be transferred to fine-scale prediction within a single unified framework. Cast3-ENS achieves leading synoptic-scale skill by broadly sampling forecast uncertainty across structurally diverse grid configurations, while Cast3-Control recovers realistic mesoscale structure through generative nudging, bridging the gap between the predictability of an ensemble mean and the physical realism of a high-resolution forecast.

The generative nudging mechanism achieves this by embedding cross-scale coupling directly into the diffusion sampling process. At every denoising step, the gradient-based correction toward the ensemble mean and the stochastic generation of fine-scale structure occur simultaneously and influence each other, so that different scales remain dynamically coupled throughout inference rather than being treated sequentially as in conventional downscaling. This coupling maintains physical consistency across scales without recourse to any external numerical model, at a computational cost limited to running a single lightweight generative model. The constraint itself adapts with lead time through the concept of effective resolution: the ensemble mean retains coherent structure only at scales where individual members agree in phase, and as members progressively diverge at longer horizons, this threshold shifts to ever larger scales. Generative nudging therefore guides only those scales at which the mean carries physically meaningful information, allowing the diffusion process to freely generate structure at finer scales.

The framework is designed to be broadly applicable. The posterior sampling mechanism is agnostic to the source of the ensemble mean: forecasts from data-driven, numerical, or hybrid systems can all be aggregated into a conditioning field. The variable set of the second stage is also decoupled from that of the ensemble, enabling the generation of quantities absent from the underlying models. This opens the possibility of tailoring the output to specific applications, such as hub-height wind for energy forecasting or surface irradiance for photovoltaic resource assessment, while inheriting dynamical constraints from the large-scale flow. The second stage can likewise be configured with additional vertical levels not present in the base ensemble.

That Cast3-Control at 12.5~km surpasses IFS-Control in Z500 ACC, despite being trained on IFS forecast fields that in principle bound its standalone skill, calls for explanation. Each ensemble member represents an independent atmospheric trajectory carrying complementary information about the atmospheric state. Aggregation extracts the shared predictable signal that no single member can isolate. Generative nudging then injects this distilled information into a physically consistent realization, effectively granting a single forecast access to the collective information content of the full ensemble. Cast3-Control therefore draws on more information than any individual member or any model trained on a single data source, explaining how it transcends a skill ceiling that would otherwise be binding.

The cubed-sphere topology underpinning Cast3 carries architectural consequences that extend beyond this study. Each of its six faces is a regular two-dimensional array compatible with standard convolutional operations, avoiding the computational overhead and optimization difficulty of graph neural networks on unstructured meshes. The six-face layout maps naturally onto a distributed computing architecture for spatial parallelism, and in principle allows the model to scale to substantially larger sizes. Recent spectral analysis has shown that GenCast exhibits persistent spectral artefacts at small scales \cite{kim2026spectral}, reflecting the difficulty of learning faithful fine-scale structure when the generative model must operate on a high-dimensional state space. The stretched cubed-sphere grid offers a natural remedy: by concentrating resolution over the target region, it reduces the effective dimensionality of the generation task by approximately a factor of four, making it substantially easier for the diffusion model to reproduce realistic fine-scale dynamics and rendering further extension to higher resolutions more tractable.

Looking ahead, the cascaded architecture provides a natural pathway for absorbing the rapidly growing archives of km-scale numerical simulations and convection-permitting reanalysis products. Each resolution stage can be independently retrained or extended without modifying the base system, and the posterior sampling mechanism ensures that improvements in either the ensemble or the high-resolution generator propagate directly to the final forecast. More broadly, the combination of cubed-sphere topology and spatial parallelism offers a scalable foundation for data-driven atmospheric modelling, in which resolution, ensemble size, and training data can be grown independently along well-defined axes without architectural modification.

\section{Method}

\subsection*{Task formulation}
We formulate the forecasting problem as sampling from the conditional distribution $p(\mathbf{x}_t | \mathbf{x}_{t-1})$. This approach aims to capture two distinct sources of uncertainty: the intrinsic stochasticity of the atmosphere, and the systematic bias arising from the inevitable imperfections in model optimization and grid discretization.

To approximate this complex high-dimensional distribution, we learn the conditional probability density function by training a probabilistic diffusion model \cite{ho2020denoising,song2020score} on massive datasets of atmospheric reanalysis. Mathematically, the diffusion model functions as a learnable transition kernel that iteratively maps a standard Gaussian prior to the target atmospheric data distribution conditioned on the previous state $\mathbf{x}_{t-1}$.

Crucially, single-model formulations often fail to fully capture epistemic uncertainties stemming from structural limitations, such as specific mesh approximations or local minima in optimization. To address this, we introduce a grid-based Monte Carlo sampling strategy to generate a diverse forecast ensemble $\{\mathbf{x}_t^{(k)}\}_{k=1}^K$. This strategy explicitly accounts for structural and optimization variances by aggregating predictions across a heterogeneous set of models. Specifically, we vary the underlying discrete geometry using different Cubed-Sphere grid configurations and diversify the model parameters via distinct random initialization seeds and training epoch snapshots. By sampling from this broad model space, we effectively marginalize over the systematic uncertainties that remain unquantifiable in single-mesh formulations, thereby providing a comprehensive representation of potential atmospheric states.

To achieve a forecast that retains the high predictive skill of an ensemble mean while preserving the realistic spectral energy of individual realizations, we formulate a controllable generation process. We introduce a super-ensemble mean, $\mathbf{\bar{x}}_{\text{ens}}$, derived from external large-scale forecasting systems, which serves as a low-frequency physical constraint.

We redefine the forecasting objective as sampling from the posterior distribution conditioned on the validity of this ensemble mean:
\begin{equation}
    p(\mathbf{x}_t \mid \mathbf{x}_{t-1}, \mathcal{S}(\mathbf{\bar{x}}_{\text{ens}}, \sigma))
\end{equation}
where $\mathcal{S}$ denotes a constraint operator governed by the super-ensemble mean $\mathbf{\bar{x}}_{\text{ens}}$ and its effective reliability scale $\sigma$. In this formulation, $\mathbf{x}_{t-1}$ provides the initial boundary condition for the stochastic dynamics, while $\mathcal{S}(\mathbf{\bar{x}}_{\text{ens}}, \sigma)$ acts as a guidance term in the generative process, restricting the solution space to physically plausible regions that align with the high-confidence modes of the super-ensemble.

\subsection*{Generative Modeling of Atmospheric Dynamics}

Atmospheric states exhibit strong temporal correlations dominated by slow-varying background fields. We model the transition probability $P(\mathbf{x}_t | \mathbf{x}_{t-1})$ using a diffusion probabilistic model targeting the residual dynamics. Let $\mathbf{x}_t$ denote the atmospheric state at time $t$. Instead of predicting $\mathbf{x}_t$ directly, our network learns the residual term $\mathbf{r}_t = \mathbf{x}_t - \mathbf{x}_{t-1}$. This subtraction strategy effectively acts as a high-pass filter, simplifying the optimization landscape by focusing the generative process solely on the evolving dynamics rather than reconstructing static global features.

We formalize the diffusion process using continuous-time Stochastic Differential Equations (SDEs). The perturbation of data into noise and the subsequent restoration are governed by a forward and a reverse-time SDE, respectively. The forward SDE diffuses the data distribution $p_{\text{data}}(\mathbf{x})$ into a prior noise distribution $p_T(\mathbf{x})$ over a continuous time variable $\tau \in [0, T]$:
\begin{equation}
    \mathrm{d}\mathbf{x} = \mathbf{f}(\mathbf{x}, \tau) \mathrm{d}\tau + g(\tau) \mathrm{d}\mathbf{w},
\end{equation}
where $\mathbf{w}$ denotes the standard Wiener process, $\mathbf{f}(\mathbf{x}, \tau)$ is the drift coefficient that controls deterministic evolution, and $g(\tau)$ is the diffusion coefficient that governs stochastic perturbations. 

We adopt the Variance Preserving (VP) SDE formulation, which specifies the drift and diffusion coefficients as:
\begin{equation}
    \mathbf{f}(\mathbf{x}, \tau) = -\frac{1}{2} \beta(\tau) \mathbf{x}, \quad g(\tau) = \sqrt{\beta(\tau)},
\end{equation}
where $\beta(\tau)$ is a time-dependent noise variance schedule. The drift term $\mathbf{f}(\mathbf{x}, \tau)$ shrinks the signal linearly with rate proportional to $\beta(\tau)$, while the diffusion term $g(\tau)$ injects isotropic Gaussian noise. This specific parametrization ensures that the marginal distribution of $\mathbf{x}(\tau)$ remains a Gaussian with unit variance if the initial data is normalized, simplifying the training dynamics.

The generative process corresponds to the reverse-time SDE, which reconstructs the atmospheric state from the noise prior:
\begin{equation}
    \mathrm{d}\mathbf{x} = \left[ \mathbf{f}(\mathbf{x}, \tau) - g(\tau)^2 \nabla_{\mathbf{x}} \log p_\tau(\mathbf{x}) \right] \mathrm{d}\tau + g(\tau) \mathrm{d}\bar{\mathbf{w}},
\end{equation}
where $\mathrm{d}\bar{\mathbf{w}}$ is the Wiener process in reverse time, and $\nabla_{\mathbf{x}} \log p_\tau(\mathbf{x})$ is the score function. Our neural network approximates this score function, allowing us to numerically solve Eq. (2) to generate physically consistent atmospheric residuals.

We discretize the continuous time domain into $N=1000$ steps. The evolution of the noise schedule $\beta(\tau)$ follows a cosine function. Specifically, the cumulative noise variance $\bar{\alpha}_\tau$ is defined as:
\begin{equation}
    \bar{\alpha}_\tau = \frac{f(\tau)}{f(0)}, \quad \text{where } f(\tau) = \cos^2\left(\frac{\tau/T + s}{1 + s} \cdot \frac{\pi}{2}\right),
\end{equation}
with a small offset $s$ to prevent singularities at $\tau=0$. The instantaneous beta schedule is then derived as $\beta(\tau) = -\frac{\mathrm{d}}{\mathrm{d}\tau} \log \bar{\alpha}_\tau$. This scheduling strategy modulates the signal decay rate smoothly, ensuring that the signal-to-noise ratio decreases gradually. Such design is critical for retaining fine-grained meteorological structures throughout the diffusion process.

\begin{figure}[h]
\centering
\includegraphics[width=\textwidth]{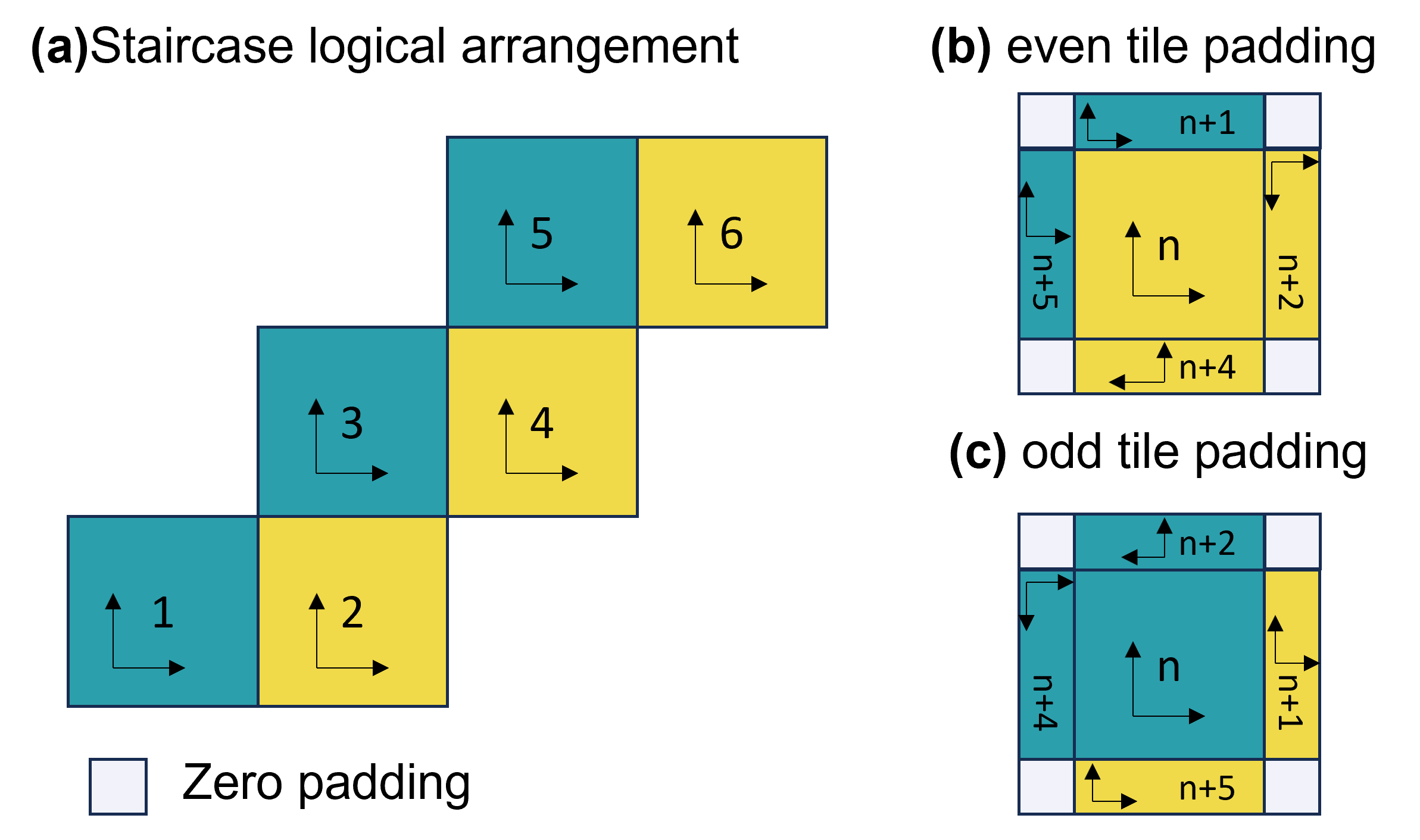}
\caption{Topology-aware padding on the Cubed-Sphere grid. \textbf{a} Staircase unfolding of the six faces with local coordinate orientations indicated by arrows. \textbf{b} Padding construction for even-numbered faces. \textbf{c} Padding construction for odd-numbered faces. For each face $n$, boundary strips from four topological neighbors ($n\!+\!1$, $n\!+\!2$, $n\!+\!4$, $n\!+\!5$ modulo 6) are geometrically transformed to match the target face's coordinate system and appended as convolution padding. Corner regions (white) are zero-filled.}
\label{fig:padding}
\end{figure}

Representing global atmospheric states requires handling spherical geometry while avoiding polar singularities. We address this by grounding our architecture in a Cubed-Sphere (CS) topology. A central challenge lies in applying standard 2D convolutions across discontinuous face boundaries, where adjacent faces possess distinct local coordinate orientations. Inspired by the domain decomposition and halo-exchange procedures employed in the FV3 dynamical core \cite{chen_lmars_2021}, we resolve this via a \textbf{topology-aware padding strategy}. As illustrated in Fig.~\ref{fig:padding}, the six faces of the Cubed-Sphere are unfolded into a logical staircase arrangement, forming three pairs of panels with alternating coordinate handedness. For each face $n$, boundary strips are extracted from its four topological neighbors ($n\!+\!1$, $n\!+\!2$, $n\!+\!4$, $n\!+\!5$ modulo 6) and geometrically transformed to align with the target face's local coordinate system before being appended as convolution padding. Even- and odd-numbered faces follow distinct adjacency and transformation rules dictated by the Cubed-Sphere topology. This mechanism ensures strictly continuous receptive fields across all face boundaries, allowing standard 2D convolution kernels to operate without introducing edge discontinuities. We implement the backbone using a CS-adapted ConvNeXtV2 \cite{woo2023convnext} in a U-shaped encoder-decoder structure. The network downsamples the spatial resolution twice to capture multi-scale interactions. We assign channel dimensions of 384, 768, and 1536 at successive stages. The corresponding block depths are set to 3, 3, and 27, respectively. This deep hierarchical configuration balances computational efficiency with the capacity to model complex turbulent flows.

We utilize \textbf{v-prediction} \cite{salimans2022progressive} as the reparameterization target to enhance training stability. The network predicts a velocity vector $\mathbf{v} \equiv \alpha_\tau \epsilon - \sigma_\tau \mathbf{x}$, where $\alpha_\tau$ and $\sigma_\tau$ are signal and noise scale factors. This approach mitigates numerical instabilities at extreme signal-to-noise ratios. We optimize the model using a weighted Mean Squared Error (MSE) loss. The objective function incorporates area weighting and variable-specific importance:
\begin{equation}
    \mathcal{L} = \mathbb{E}_{\tau, \mathbf{x}, \epsilon} \left[ \sum_{k} \lambda_k \sum_{i} A_i \left\| \mathbf{v}_{\text{pred}}^{(k, i)} - \mathbf{v}_{\text{target}}^{(k, i)} \right\|^2 \right].
\end{equation}
Here, $A_i$ represents the normalized area weight for grid cell $i$. The term $\lambda_k$ denotes the weight for physical variable $k$. We apply the same weighting following recent protocols \cite{lam_learning_2023}. The training is conducted on 48 Sugon BW100 accelerators (6 nodes, 8 accelerators per node), each equipped with 64\,GB of HBM (1.8\,TB/s bandwidth) and delivering 74\,TFLOPS of FP32 peak throughput. Accelerators within each node are connected through a high-speed inter-accelerator interconnect. The software environment uses a GPGPU programming framework. We perform 180,000 iterations in total over the course of approximately 21 days. The learning rate increases linearly to $2.5 \times 10^{-4}$ during a 3,000-step warmup phase and subsequently decays to zero following a cosine schedule. We employ the AdamW optimizer with a weight decay of 0.1.

Upon the completion of training, efficient sampling becomes critical for operational deployment. Standard SDE solvers are often computationally prohibitive for real-time forecasting. To accelerate inference, we adopt the Denoising Diffusion Implicit Models (DDIM) \cite{song2020denoising} approach. This method reformulates the reverse process as a deterministic Ordinary Differential Equation (ODE), allowing us to bypass redundant steps. We generate forecast ensembles using a 20-step sampling trajectory. This configuration strikes an optimal balance between generation speed and the accuracy of the atmospheric energy spectrum.

\subsection*{Multi-Grid Ensemble for Uncertainty Quantification}

Global atmospheric models must navigate the trade-off between regional resolution and computational cost. Recent data-driven approaches often mitigate this burden using patch-based embeddings or latent representations \cite{bi2023accurate,chen2023fuxi,kossaifi2026demystifying}. However, such spatial compression can disrupt continuous physical dependencies. This degradation frequently manifests as unphysical artifacts in the Power Spectral Density (PSD). We address the resolution constraint while preserving physical topology using a stretched-grid configuration. This approach dynamically refines resolution over specific target regions. It maintains coarser spacing elsewhere to preserve efficiency. Our implementation achieves a local resolution of 25 km over areas of interest. The global average resolution remains approximately 50 km.

Discretization choices introduce structural uncertainty into forecasts. We adopt a superensemble framework to quantify this effect \cite{krishnamurti_review_2016}. Distinct grid orientations emphasize different geographical features. Consequently, each configuration yields unique error characteristics. We capture this spread by training multiple forecast models, each operating on a different grid topology. Marginalizing over these grid-dependent errors allows the ensemble to filter out discretization noise. This process effectively isolates the predictable large-scale signal.

We generate the stretched grids using the Schmidt transformation \cite{schmidt_variable_1977}. The transformation concentrates grid cells near the target coordinates and expands cells in the antipodal region. This process preserves the orthogonality and smoothness of the original topology. We select a stretching factor that yields the target 25 km local resolution.

We construct an ensemble comprising four distinct stretched-grid configurations and one uniform baseline (Table~\ref{tab:grid_configs}). Each stretched grid centers refinement over a different continental region. We train separate models independently on each grid. This design allows the ensemble to represent physical uncertainty alongside the structural uncertainty derived from spatial discretization.

\begin{table}[h]
\centering
\caption{Multi-grid ensemble configurations. Stretched grids achieve 25 km local resolution with a $\sim$50 km global mean. The uniform grid maintains constant spacing.}
\label{tab:grid_configs}
\begin{tabular}{lcccc}
\toprule
\textbf{Grid ID} & \textbf{Focus Region} & \textbf{Local Resolution} & \textbf{Global Mean} & \textbf{Grid Type} \\
\midrule
Grid-CN & China & 25 km & 50 km & Stretched C192 \\
Grid-US & United States & 25 km & 50 km & Stretched C192 \\
Grid-EU & Europe & 25 km & 50 km & Stretched C192 \\
Grid-Unif & Global & 50 km & 50 km & Uniform C192 \\
Grid-Unif & Global & 100 km & 100 km & Uniform C96 \\
\bottomrule
\end{tabular}
\end{table}

\subsection*{Cascaded Resolution Refinement}

Training high-resolution forecast models from scratch is computationally prohibitive and data-intensive. Inspired by cascaded generation strategies in image synthesis \cite{yu2025pixeldit}, we adopt a progressive refinement framework that leverages pretrained coarse-resolution models as condition. This design decomposes the forecasting task across spatial scales: the coarse model captures large-scale dynamical features, while a lightweight refiner concentrates on recovering fine-scale details.

The cascaded architecture requires strict spatial alignment between successive resolution stages. We construct stretched-grid pairs that share identical focus centers and stretching parameters but differ by a factor of two in resolution. Concretely, we pair the Grid-CN configuration (C192, 25\,km local resolution) with a C384 counterpart centered on the same region, achieving 12.5\,km local resolution and approximately 25\,km global mean spacing. Because both grids derive from the same Schmidt transformation parameters, each coarse grid cell maps exactly onto a $2 \times 2$ block of fine grid cells. This bijective correspondence eliminates the need for spatial interpolation between stages.

During training, the pretrained C192 model is frozen and serves as a guidance generator. Given a noisy high-resolution state $\mathbf{x}_t$ on the C384 grid, we obtain its coarse counterpart $\mathbf{x}_t^{\mathrm{low}}$ via stride-2 spatial slicing, which preserves the noise variance by construction. The frozen model produces a denoised low-resolution prediction $\hat{\mathbf{x}}_0^{\mathrm{low}}$, which is then upsampled to the C384 grid through learned transposed convolutions to yield a guidance signal $\mathbf{g}$. The refiner receives the concatenation of the noisy state $\mathbf{x}_t$, the meteorological conditions $\mathbf{c}$, and the guidance $\mathbf{g}$ as input.

The refiner adopts a shallow U-Net architecture with two stages. The encoder operates at channel dimension 256, followed by a single downsampling step to a 512-dimensional bottleneck. Skip connections link the encoder to a symmetric decoder via channel-wise concatenation. We use a cosine-scheduled diffusion process conditioned on sinusoidal time embeddings. The refiner contains substantially fewer parameters than the base model, as it is only required to learn the residual high-frequency structure that the coarse model cannot resolve.

We train the refiner using forecast fields from the IFS HRES as supervision (see Data). This cascaded formulation is naturally extensible: by constructing additional stretched-grid pairs (e.g., C384$\to$C768), subsequent refiners can progressively approach kilometre-scale resolution without retraining lower-resolution stages. Each model in the cascade retains independent forecast utility, and the marginal training cost decreases at each successive stage owing to the increasingly localized nature of the residual learning task.

\subsection*{Generative Nudging}

We introduce a generative nudging framework to produce a control forecast state $\mathbf{x}_t$. This approach combines the high large-scale skill of the ensemble mean with the realistic small-scale variability. The generation process is constrained by two factors: the previous high-resolution state $\mathbf{x}_{t-1}$ and the reliable scales of the current ensemble mean $\bar{\mathbf{x}}_t$.

We first determine the effective spatial scale of the ensemble mean. This scale defines the resolution limit where $\bar{\mathbf{x}}_t$ retains physical validity. We analyze the power spectral density (PSD) of the ensemble mean relative to ERA5 reanalysis. As forecast lead time increases, the energy of the ensemble mean decays at higher wavenumbers. We define the \emph{effective resolution} as the spatial wavelength at which the spherical harmonic power spectral density of $\bar{\mathbf{x}}_t$ falls below half that of ERA5 reanalysis. At short lead times, constructive interference among ensemble members preserves energy down to relatively fine scales; as the forecast horizon extends, phase errors among members grow and their destructive interference progressively erodes energy at higher wavenumbers. Table~\ref{tab:effective_resolution} summarizes the effective resolution as a function of forecast lead time. At Day~1, the ensemble mean retains skill down to wavelengths of approximately 790\,km (wavenumber $\sim$50, equivalent to $\sim$C25 resolution). By Day~5, this threshold coarsens to $\sim$1625\,km ($\sim$C12), and by Day~10 only structures larger than $\sim$3260\,km ($\sim$C6) remain reliable. We implement the low-pass filter $\mathcal{S}$ as a spatial mean-pooling operation configured at each lead time according to these empirically determined cutoffs. This operator averages out sub-grid variability below the effective scale. Consequently, the constraint field $\mathbf{c}_t = \mathcal{S}(\bar{\mathbf{x}}_t)$ retains only information the ensemble can credibly resolve.

\begin{table}[h]
\centering
\caption{Effective resolution of the ensemble mean as a function of forecast lead time. The effective wavelength corresponds to the scale at which the ratio of ensemble-mean spectral energy to ERA5 spectral energy drops below 0.5.}
\label{tab:effective_resolution}
\begin{tabular}{cccc}
\toprule
\textbf{Lead Time (days)} & \textbf{Eff.\ Wavelength (km)} & \textbf{Wavenumber} & \textbf{Equiv.\ Resolution} \\
\midrule
1  & 792   & 50.5 & $\sim$C25 \\
2  & 946   & 42.3 & $\sim$C21 \\
3  & 1099  & 36.4 & $\sim$C18 \\
4  & 1309  & 30.6 & $\sim$C15 \\
5  & 1625  & 24.6 & $\sim$C12 \\
6  & 1874  & 21.4 & $\sim$C11 \\
7  & 2300  & 17.4 & $\sim$C9  \\
8  & 2542  & 15.8 & $\sim$C8  \\
9  & 3065  & 13.1 & $\sim$C7  \\
10 & 3259  & 12.3 & $\sim$C6  \\
\bottomrule
\end{tabular}
\end{table}

We enforce alignment with this constraint by sampling from the conditional posterior $p(\mathbf{x}_t \mid \mathbf{x}_{t-1}, \mathbf{c}_t)$. By applying Bayes' rule, the score function decomposes into an unconditional prior term and a likelihood guidance term \cite{chung2022diffusion}:
\begin{equation}
\nabla_{\mathbf{z}_\tau} \log p(\mathbf{z}_\tau \mid \mathbf{x}_{t-1}, \mathbf{c}_t) = \nabla_{\mathbf{z}_\tau} \log p_\theta(\mathbf{z}_\tau \mid \mathbf{x}_{t-1}) + \nabla_{\mathbf{z}_\tau} \log p(\mathbf{c}_t \mid \mathbf{z}_\tau)
\end{equation}
where $\mathbf{z}_\tau$ denotes the noisy latent state at diffusion step $\tau$.

For the guidance term $\nabla_{\mathbf{z}_\tau} \log p(\mathbf{c}_t \mid \mathbf{z}_\tau)$, we adopt the approximation proposed in Diffusion Posterior Sampling (DPS) \cite{chung2022diffusion}. We model the deviation from the constraint as Gaussian noise. DPS demonstrates that the gradient of the log-likelihood can be approximated using the estimated clean state $\hat{\mathbf{x}}_t(\mathbf{z}_\tau)$, derived via Tweedie's formula. This yields a tractable gradient update based on the scale-aware $L_2$ distance:
\begin{equation}
\nabla_{\mathbf{z}_\tau} \log p(\mathbf{c}_t \mid \mathbf{z}_\tau) \simeq -\zeta \nabla_{\mathbf{z}_\tau} \| \mathbf{c}_t - \mathcal{S}(\hat{\mathbf{x}}_t(\mathbf{z}_\tau)) \|^2_2
\end{equation}
Here, $\zeta$ is a scalar hyperparameter controlling the guidance strength. This formulation ensures that the generated trajectory is iteratively ``nudged'' to align with the reliable large-scale flow $\mathbf{c}_t$. Simultaneously, the learned diffusion foreacast model naturally reconstructs the high-frequency details undefined by the ensemble mean.

\subsection*{Data}

The base models are trained on ERA5 reanalysis \cite{hersbach2020era5} produced by the European Centre for Medium-Range Weather Forecasts (ECMWF). The modelling target comprises 69 atmospheric variables on a $721 \times 1440$ latitude-longitude grid at 0.25-degree resolution. Five upper-air fields are extracted at 13 pressure levels (50, 100, 150, 200, 250, 300, 400, 500, 600, 700, 850, 925, and 1000\,hPa): geopotential (Z), specific humidity (Q), temperature (T), zonal wind (U), and meridional wind (V). Together, these fields characterize the thermodynamic and kinematic state of the atmosphere from the lower stratosphere to the planetary boundary layer. Four surface variables complement the upper-air representation: 2-metre temperature (T2M), the two horizontal components of 10-metre wind (U10, V10), and mean sea-level pressure (MSL). These quantities encode the near-surface boundary conditions that critically govern short-range weather evolution. We further supply static geographical covariates, including the land-sea mask, terrain elevation, and normalized spatial coordinates, to enable the model to learn regionally dependent dynamical patterns.

We use ERA5 data spanning 1979 to 2020 as the training corpus. Because the diffusion framework involves no hyperparameter selection that requires early stopping or model selection on held-out samples, we do not reserve a separate validation set. For each grid configuration listed in Table~\ref{tab:grid_configs}, the latitude-longitude fields are remapped onto the corresponding Cubed-Sphere mesh using the appropriate Schmidt transformation, producing five parallel training datasets that share identical physical content but differ in spatial discretization.

Training the cascaded high-resolution refiner (C384) requires supervision at 12.5\,km local resolution. The volume of available HRES analysis data at this scale is limited. To overcome this constraint, we supplement the training corpus with operational forecast fields from the ECMWF Integrated Forecasting System (IFS). Each IFS initialization produces a 10-day deterministic trajectory, effectively yielding an order-of-magnitude expansion of the available training samples: the forecast archive can be viewed as approximately ten years of parallel atmospheric realizations. This strategy allows the refiner to learn the resolved dynamical modes of the IFS at high resolution. The variable set is kept identical to that of the ERA5 training to ensure consistency across the cascaded pipeline. We note that the skill of the current high-resolution refiner remains limited relative to the base models, for two principal reasons. First, the refiner is trained on IFS forecast fields rather than on analysis or reanalysis products. Forecast fields inevitably carry model-inherent biases and error growth that are absent from objectively constrained analyses; consequently, the refiner learns to reproduce the dynamical modes of the IFS forecasting system, including its systematic deficiencies. Second, the available training corpus spans only approximately one year of operational forecasts. This restricted sample size provides insufficient coverage of the full climatological variability, limiting the refiner's ability to generalize across the diverse range of synoptic regimes present in operational conditions.

For station-based verification over China, we use quality-controlled hourly observations from 2,168 national meteorological stations operated by the China Meteorological Administration (CMA). These stations are distributed across the Chinese mainland and provide independent ground-truth measurements of surface variables including 2-metre temperature and 10-metre wind components.

\subsection*{Spatial Parallelism via Distributed Halo Exchange}

Standard deep learning parallelism strategies such as Distributed Data Parallel (DDP) replicate the full model on each accelerator and partition only the batch dimension. While this approach scales training throughput, it does not reduce the per-device memory footprint of individual samples. For high-resolution atmospheric fields on Cubed-Sphere grids, a single training sample can exceed the memory capacity of a single accelerator, rendering conventional DDP infeasible. Pipeline parallelism alleviates the parameter memory constraint by sharding the model across devices, but introduces sequential dependencies that limit wall-clock efficiency.

We draw inspiration from domain decomposition techniques widely employed in operational NWP systems \cite{putman_cross-platform_2005}, where the global computational domain is partitioned into spatial tiles, each assigned to a separate processor. Adjacent tiles exchange boundary information at every integration step to maintain physical continuity across partition boundaries. We adapt this paradigm to the Cubed-Sphere neural network by exploiting its natural six-face topology: each of the six faces is assigned to a dedicated GPU, and face-boundary data are communicated through a differentiable halo exchange protocol.

At every convolutional layer, each GPU extracts narrow boundary strips (halo regions) from its local feature map and exchanges them with topological neighbors via collective communication. The received halos are geometrically transformed according to the Cubed-Sphere adjacency rules (transposition and reflection) and appended as padding before the convolution kernel is applied. To preserve end-to-end gradient flow across face boundaries, the halo exchange is implemented as a differentiable operator: the forward pass uses \texttt{all\_gather} to distribute boundary data, while the backward pass employs \texttt{reduce\_scatter}, its exact mathematical adjoint, to route gradients back to the originating faces. Only the boundary strips are communicated rather than entire face tensors, reducing inter-device traffic by approximately a factor of six.

This spatial decomposition reduces per-device memory consumption to approximately one-sixth of the single-GPU baseline, enabling training at resolutions that would otherwise exceed device memory limits. Training throughput scales by a factor of six when using six GPUs, at the cost of a proportional increase in the total number of accelerators. Efficient operation requires high-bandwidth inter-device interconnects (e.g., NVLink) to prevent halo communication from becoming a bottleneck; for this reason, all training experiments in this work use the standard single-GPU configuration, and spatial parallelism is deployed exclusively at inference time.

Table~\ref{tab:inference_speed} reports inference wall-clock times for a 10-day forecast using the stretched C192 grid with 20-step DDIM sampling. On a single NVIDIA RTX 5090, inference requires approximately 260\,s. Distributing the six faces across six RTX 5090 GPUs reduces the per-step latency to yield a total forecast time of approximately 60\,s, a four-fold acceleration. For reference, GenCast \cite{price_probabilistic_2025} requires approximately 35 minutes for a 10-day forecast on an A100 GPU with a pre-compiled computational graph, and approximately 50 minutes with cold-start compilation overhead.

\begin{table}[h]
\centering
\caption{Inference wall-clock time for a 10-day deterministic forecast (20-step DDIM, stretched C192 grid). GenCast timing is measured on a single A100 GPU for comparison.}
\label{tab:inference_speed}
\begin{tabular}{lccc}
\toprule
\textbf{Configuration} & \textbf{Hardware} & \textbf{Time per Step} & \textbf{Total (10 days)} \\
\midrule
Ours (single GPU)           & 1$\times$ RTX 5090 & 13.0\,s & 260\,s \\
Ours (spatial-parallel)     & 6$\times$ RTX 5090 & 3.0\,s  & 60\,s  \\
GenCast (compiled)          & 1$\times$ A100     & ---     & $\sim$35\,min \\
GenCast (cold start)        & 1$\times$ A100     & ---     & $\sim$50\,min \\
\bottomrule
\end{tabular}
\end{table}

\subsection*{Verification Metrics}

We evaluate large-scale forecast skill using the latitude-weighted Anomaly Correlation Coefficient (ACC). For a given variable field, the ACC is defined as:
\begin{equation}
\mathrm{ACC} = \frac{\sum_{i} w_i \, (f_i - c_i)(a_i - c_i)}{\sqrt{\sum_{i} w_i \, (f_i - c_i)^2} \sqrt{\sum_{i} w_i \, (a_i - c_i)^2}}
\end{equation}
where $f_i$ and $a_i$ denote the forecast and analysis values at grid cell $i$, $c_i$ is the climatological mean, and $w_i = \cos(\phi_i)$ is the area weight corresponding to the latitude $\phi_i$ of grid cell $i$.

To assess the fidelity of resolved atmospheric dynamics across spatial scales, we compute the kinetic energy spectrum at 200\,hPa. The horizontal wind field $\mathbf{u} = (u, v)$ is decomposed into its rotational and divergent components via Helmholtz decomposition:
\begin{equation}
\mathbf{u} = \nabla \times (\psi \, \hat{\mathbf{k}}) + \nabla \chi
\end{equation}
where $\psi$ is the streamfunction and $\chi$ is the velocity potential. Expanding both fields in spherical harmonics, the total kinetic energy spectrum as a function of wavenumber $n$ is given by:
\begin{equation}
E(n) = \frac{n(n+1)}{4a^2} \left( |\psi_n|^2 + |\chi_n|^2 \right)
\end{equation}
where $a$ is the Earth's radius, and $\psi_n$ and $\chi_n$ are the spherical harmonic coefficients of the streamfunction and velocity potential at total wavenumber $n$, respectively.

We complement gridded verification with point-wise evaluation at observing station locations. The root-mean-square error is computed between forecasted and observed values at each station and averaged across all verification cases:
\begin{equation}
\mathrm{RMSE} = \frac{1}{N} \sum_{j=1}^{N} \sqrt{ \frac{1}{M} \sum_{k=1}^{M} \left( f_{j,k} - o_{j,k} \right)^2 }
\end{equation}
where $M$ is the number of verification cases, $N$ is the number of stations, $f_{j,k}$ and $o_{j,k}$ denote the forecast and observed values at station $k$ for case $j$.

Tropical cyclones are tracked by identifying and following minima in the mean sea-level pressure (MSLP) field, following the methodology of Pangu-Weather \cite{bi2023accurate}. Starting from a known initial cyclone position, we apply a 6-hourly iterative procedure in which local MSLP minima are detected subject to physically motivated criteria: in the Northern Hemisphere, a candidate must be collocated with an 850\,hPa relative vorticity maximum exceeding $5 \times 10^{-5}$\,s$^{-1}$ within a 278\,km search radius; in the Southern Hemisphere, the corresponding criterion requires an 850\,hPa relative vorticity minimum below $-5 \times 10^{-5}$\,s$^{-1}$ within the same radius. To maintain track continuity during extratropical transition, we alternatively identify the maximum 200--850\,hPa thickness within 278\,km. For landfalling systems, the search criterion relaxes to a 10-metre wind speed exceeding 8\,m\,s$^{-1}$ within 278\,km.

\subsection*{Model Configuration}

We benchmark our system against both data-driven and physics-based operational baselines. For the data-driven baseline, we adopt GenCast \cite{price_probabilistic_2025}. Inference with the 0.25-degree GenCast model is computationally prohibitive for large ensemble sizes; we therefore use the 1-degree GenCast to generate 50-member ensembles for large-scale skill evaluation (ACC), while employing the 0.25-degree GenCast for single-member deterministic forecasts in station-based and tropical cyclone case evaluations. For the physics-based baselines, we use the ECMWF IFS high-resolution control forecast (HRES, 9\,km) and the IFS ensemble prediction system (ENS, 51 members at 25\,km). ACC is computed against the IFS ENS, while station-based verification and tropical cyclone case studies use the IFS HRES deterministic forecast.

Our system generates ensemble forecasts by aggregating members across three grid configurations: Grid-CN (China focus), Grid-US (United States focus), and uniform C96, yielding a total of 144 ensemble members for ACC evaluation. Owing to the computational efficiency of our architecture, the 144-member ensemble remains faster to produce than the 50-member GenCast ensemble. The uniform C192 and Grid-EU configurations are excluded from the current evaluation due to insufficient training at the time of submission; incorporating these additional grids is expected to further improve ensemble skill.

In the generative nudging stage, the 144-member ensemble mean serves as the large-scale constraint to guide two deterministic forecast models: a Grid-CN C192 model (25\,km over China, trained on ERA5) and a Grid-CN C384 model (12.5\,km over China, trained on IFS forecast fields). As noted in the Data section, the limited volume and inherent biases of the IFS forecast training data constrain the skill of the C384 refiner. We therefore adopt the ERA5-trained C192 Grid-CN model as the unified configuration for station-based evaluation and tropical cyclone case studies, where local forecast accuracy is critical.

\backmatter




\newpage
\bmhead{Acknowledgements}
This research was supported by the Strategic Priority Research Program of the Chinese Academy of Sciences (Grant No. XDA0500000), the National Natural Science Foundation of China (42275174), and the National Key R\&D Program of China (2025YFF0517203, 2022YFF0802001). The authors gratefully acknowledge the OneScience platform for providing comprehensive computational infrastructure, environment configuration, and support for model training and inference throughout this work. We thank the HPC Department of Sugon for technical support and TianJi Weather Science and Technology Company for collaborative contributions. Real-time operational forecasting of Cast3 is hosted on the Earth System Science Numerical Simulator Facility (EarthLab), whose sustained support is gratefully acknowledged.

\newpage
\bibliography{sn-bibliography}

\end{document}